\documentclass[10pt,preprint2]{aastex}
\usepackage[left=2.0cm,top=2.0cm,right=2cm]{geometry}

\usepackage{amsmath}                % American Mathematical Society package
\usepackage{amsfonts}               % American Mathematical Society fonts
\usepackage{amssymb}                % American Mathematical Society symbol
\usepackage{epsfig}                 % EPS figures

\newcommand{\cm}{{~\rm cm}}
\newcommand{\s}{{~\rm s}}
\newcommand{\km}{{~\rm km}}

\newcommand{\erg}{{~\rm erg}}

\newcommand{\days}{{~\rm days}}

\begin{document}

\title{Powering the Second 2012 Outburst of SN 2009ip by Repeating Binary Interaction}
\author
{Amit Kashi\thanks{Email: kashia@physics.unlv.edu}, 
Noam Soker\thanks{Email: soker@physics.technion.ac.il},
and
Nitsan Moskovitz$^2$
\vspace{0.3cm}\\
\small{
$^1$ Department of Physics \& Astronomy, University of Nevada, Las Vegas, 4505 S. Maryland Pkwy, Las Vegas, NV, 89154-4002, USA \\
$^2$ Department of Physics, Technion -- Israel Institute of Technology, Haifa 32000, Israel
}
}

\begin{abstract}
We propose that the major 2012 outburst of the supernova impostor SN~2009ip was powered by an extended and repeated
interaction between the Luminous Blue Variable (LBV) and a more compact companion.
Motivated by the recent analysis of \cite{Marguttietal2013} of ejected clumps and shells we consider two scenarios.
In both scenarios the major 2012b outburst with total (radiated + kinetic) energy of $\sim 5 \times 10^{49} \erg$ was powered by accretion of $\sim 2$--$5 ~\rm{M_\odot}$
onto the companion during a periastron passage (the first passage) of the binary system approximately 20 days before the
observed maximum of the light curve.
In the first scenario, the surviving companion scenario, the companion was not destructed and still exists in the system after the outburst.
It ejected partial shells (or collimated outflows or clumps) for two consecutive periastron passages after the major one.
The orbital period was reduced from $\sim 38$ days to $\sim 25$ days as a result of the mass
transfer process that took place during the first periastron passage.
In the second scenario, the merger scenario, some partial shells/clumps were ejected also in a second periastron passage that
took place $\sim 20$ days after the first one.
After this second periastron passage the companion dived too deep into the LBV envelope to launch more outflows,
and merged with the LBV.
\end{abstract}


% ====================
\section{Introduction}
\label{sec:introduction}
% ====================

Luminous Blue Variable (LBV) is a short lasting evolutionary stage in the life of massive stars,
with a typical duration of $\sim 10^4$ years.
During this stage massive stars are prone to instabilities which might lead to giant outbursts where a significant fraction
of the stellar envelope is ejected (e.g. \citealt{Stahl1989}; \citealt{HumphreysDavidson1994}; \citealt{SmithOwocki2006}; \citealt{HarpazSoker2009}).
A companion star in an eccentric orbit may be a trigger for such an outburst, e.g., by the tidal forces it exerts on the unstable LBV
envelope when the two stars approach each other (e.g, \citealt{KashiSoker2010a}; {Kashi2010}).
Such LBV outbursts are often confused with Supernovae (SN), and hence are termed SN impostors (e.g. \citealt{Kochanek2012}; \citealt{SmithOwocki2006},\citeyear{Smithetal2011}; \citealt{DavidsonHumphreys2012} and references therein).

The SN impostor SN~2009ip, located in the spiral galaxy NGC~7259, is one of those SN impostors that are of LBV origin \citep{Berger2009}.
This LBV showed a series of outbursts, starting in 2009 \citep{Maza2009}.
SN~2009ip maintained an unstable state leading to a series of outbursts (e.g., \citealt{Drake2012}; \citealt{Levesque2012}; \citealt{Mauerhan2013}; \citealt{Pastorello2012}), resulting in an increase by 3--4 magnitudes in the V band in September 2011 and August 2012 (outburst 2012a), and by $\sim 7$ magnitudes in September 2012 (outburst 2012b).
\cite{Pastorello2012} calculated the bolometric luminosity of the outburst and found the peak of the 2012b outburst to have $L_{\rm p} = 8 \times 10^{42} \erg \s^{-1}$.
More recently, \cite{Marguttietal2013} presented observations of the outburst ranging throughout the entire spectrum, and found
that $L_{\rm p} = 1.2 \times 10^{43} \erg \s^{-1}$.
The bolometric energy radiated during the outbursts were found to be $E_{{\rm rad,}a}=(1.5 \pm 0.4) \times 10^{48} \erg$ for the 2012a outburst,
and $E_{{\rm rad,}b}=(3.2 \pm 0.3) \times 10^{49} \erg$ for the 2012b outburst \citep{Marguttietal2013}.
For an ejecta mass of $\sim 0.5 ~\rm{M_\odot}$ \cite{Marguttietal2013} estimated the total energy to be $E_{{\rm tot}}\sim 10^{50} \erg$.

\cite{Mauerhan2013} (see also \citealt{SmithMauerhan2012}) noticed the high velocity, up to $\sim 13\,000 \km \s^{-1}$,
of the gas ejected in the 2012a outburst and suggested that most of the energy radiated in the large 2012b peak came
from the kinetic energy of the material ejected during the 2012a outburst.
In the scenario proposed by \cite{Mauerhan2013}, the 2012a was a SN explosion that terminated the existence of the star.
The luminous and energetic 2012b outburst was attributed to the collision of fast SN ejecta from the 2012a outburst with
a slower shell that was ejected earlier \citep{Mauerhan2013, Prieto2013}.
Another scenario that is based on terminal stellar explosion is the dual-shock Quark-Novae proposed by \cite{Ouyedetal2013}.
The 2010a outburst is attributed to a regular SN explosion, and the 2012b outburst to a Quark-Nova explosion.

\cite{SokerKashi2013} compared the 2012a and 2012b outbursts of SN2009ip with the outburst of the intermediate luminosity optical transient
(ILOT) V838~Mon, whose outburst was found to be composed of three shell-ejection episodes \citep{Tylenda2005}.
The ejection of separate shells for the 2012a and 2012b outbursts supports the binary scenario proposed by \cite{SokerKashi2013},
and also support a single star repeated core instability scenario mentioned by \cite{Pastorello2012}.

The 2012b brightening was attributed by \cite{Marguttietal2013} to an explosive shock breaking coming from an interaction between explosive
ejection of the LBV envelope taking place $\sim 20$--$24$ days before the 2012b peak, to shells of material ejected during the 2012a eruption.
The results of \cite{Marguttietal2013} put into question \cite{Mauerhan2013} scenario.
The reason, as noted by \cite{Marguttietal2013}, is that the photosphere expansion velocity of $\sim 4500 \km \s^{-1}$
during the 2012b outburst shows that whatever gas accelerated the photosphere originated long after the peak of the 2012a event.
Namely, the gas was ejected long after the star has ceased to exist according to the scenario proposed by \cite{Mauerhan2013}.

\cite{Foley2011} suggested that the ZAMS mass of the erupting star was $M_1 \geq 60 ~\rm{M_\odot}$, assuming it was a non-rotating LBV.
Assuming a rotating LBV at 40 per cent critical velocity, \cite{Marguttietal2013} gave an estimate of $M_1 = 45$--$85 ~\rm{M_\odot}$.
Based on some similarities between the 2012 outbursts of SN2009ip and the light curve of the ILOT event V838~Mon, \cite{SokerKashi2013}
suggested that SN 2009ip was a massive binary system with an LBV of $M_1=60$--$100 ~\rm{M_\odot}$ and a main-sequence companion of
$M_2=0.2$--$0.5 M_1$ on an eccentric orbit.

Similarly to their model for the triggering of the peaks in the the $19^{\rm th}$ century Great Eruption of the LBV $\eta$ Carinae (e.g., \citealt{DavidsonHumphreys2012} and references therein),
\cite{SokerKashi2013} suggested that the peaks in the light curve of SN2009ip since 2009 were triggered during periastron passages of the binary system.
They then speculate that the very energetic 2012b outburst was a final merger process of the two stars.
In light of the high quality data obtained by \cite{Marguttietal2013}, and their analysis of SN2009ip in 2012, we re-examine here the assumption of
a final merger, and check whether it is possible that the companion survived the violent 2012b major outburst, but on a shorter period orbit.

The analysis by \cite{Marguttietal2013} emphasizes the two re-brightening peaks in the declining light-curve of the 2012b outburst, and the
appearance of absorption features at different velocities and times.
Based on these we consider two scenarios.
In section \ref{sec:peaks} we discuss the first scenario, the surviving companion scenario, in which the companion survived the 2012b outburst.
We discuss the two re-brightening peaks and raise the possibility that they are the result of a repeating binary interaction near periastron passages,
similar to our model for $\eta$ Carinae \citep{KashiSoker2010a}.
The orbital phase was shortened from $\sim 38 \days$ to $\sim 25 \days$ as a result of mass accretion.
The orbital evolution of the surviving companion scenario following the mass transfer from the LBV primary to the companion is discussed in section \ref{sec:transfer}.

The second scenario considers a terminal binary merger, but only after the system had experienced a second periastron passage after the major one.
As in the surviving companion scenario the major interaction that powered the 2012b outburst was powered by
mass accretion that shorten the orbital period. But in the merger scenario the orbit was shortened even more, and the second periastron passage occurred
$\sim 20 \days$ after the first (major) periastron passage.
After the second periastron passage the companion dived too deep into the envelope to further eject material.
In section \ref{sec:merger} we discuss the merger scenario and how ejection of collimated outflows and/or clumps during the two
periastron passages can account for the absorption features presented by \cite{Marguttietal2013}.
We summarize and discuss our main results in section \ref{sec:summary}.

% ==========================================================
\section{The Peaks During Declining}
\label{sec:peaks}
% ==========================================================

\begin{figure}[!t]
\includegraphics[trim=1.0cm 0.9cm 0cm 0.1cm, clip=true, width=1.09\columnwidth]{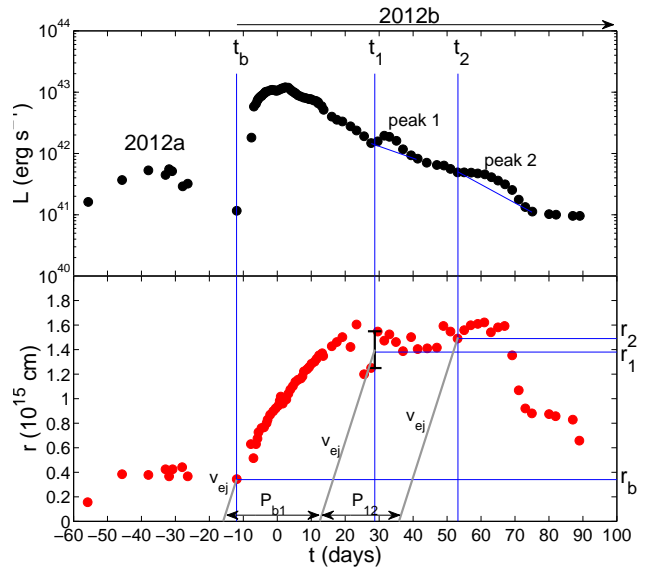}
\caption{The bolometric light curve (upper panel) and radius (lower panel) of SN 2009ip from \cite{Marguttietal2013}.
We mark times $t_b$, $t_1$ and $t_2$ and corresponding radii $r_b$, $r_1$ and $r_2$ for the rise of the 2012b event, and the two re-brightening
peaks, respectively.
The rise of each peak starts when the ejecta reaches the photosphere.
The gray lines in the lower panel represent the propagation of the ejected shells
from the center to their respective radii according to the surviving companion scenario, assuming an ejection velocity of $v_{\rm ej} = 10^4 \km \s^{-1}$.
On the formation of $\sim 10^4 \km \s^{-1}$ ejecta near periastron passages in the binary  model see \cite{TsebrenkoSoker2013}.
We mark the periods $P_{b1}$ (equation \ref{eq:Pb1}) and $P_{12}$ (equation \ref{eq:P12}) which we deduce from the light curve and the ejection velocity.}
\label{fig:margutti_lightcurve}
\end{figure}

The bolometric light curve of SN~2009ip shows two peaks during the decline phase of the 2012b outburst
(\citealt{Marguttietal2013}; see also \citealt{Martinetal2013a}; \citealt{Hambsch2012}; \citealt{Martinetal2012} and \citealt{Margutti2012ATel}).
In Fig. \ref{fig:margutti_lightcurve} we reproduce the light curve and photospheric radius from fig. 11 of \cite{Marguttietal2013},
and mark there the starting times of the the 2012b outburst, the first peak, and the second peak,
as $t_b=-12~{\rm day}$, $t_1=28~{\rm day}$, and $t_2=53~{\rm day}$, respectively.
We note that there are also other peaks close to maximum.
We focus on the peaks during the decline because the peaks near maximum are probably not directly related to the binary period,
and may be a result of fluctuations in the accretion rate that lead
to fluctuations in the light curve.
Alternatively, there might be different clumps that result from jets interacting with a 
circumstellar matter (CSM; \citealt{TsebrenkoSoker2013}) that could cause fluctuations in the light curve. 
We here look for clear departures from a smooth light curve that can hint at a binary period,
and find these in the decline.
A more careful analysis by \cite{Martinetal2013a} reveal a dominant time scale of 24 days, similar to what we use in our modeling.
We assume that each of the two peaks during the decline is activated by the collision of ejecta from the interacting binary system.
In the surviving companion scenario discussed in this and the next section we assume that the ejecta that powered peak 1 and peak 2 were
ejected in two consecutive periastron passages that followed the major periastron passage, respectively.
The velocities of the ejecta are taken to be $v_{{\rm ej},b}$, $v_{{\rm ej},1}$, and $v_{{\rm ej},2}$, respectively.
We also assume that $v_{{\rm ej},b} \simeq v_{{\rm ej},1} \simeq  v_{{\rm ej},2} = v_{\rm ej}$,
and that the rise of each peak starts when the ejecta reaches the photosphere.
Note that the ejecta that powered peak 1 and peak 2 are not full shells, but rather clumps of collimated outflows.
% FFFFFFFFFFFFFFFFFFFFFFFFFFFFFFFFFFFFFFFFFFFFFF
% FFFFFFFFFFFFFFFFFFFFFFFFFFFFFFFFFFFFFFFFFFFFFF

The photospheric radii at the onset of the rise to each peak are $r_b=3.4 \times10^{14} \cm$, $r_1 \simeq 1.4\times10^{15} \cm$, and $r_2 \simeq 1.5 \times10^{15} \cm$, respectively.
We estimate the extra energy in the peaks by drawing a straight line below the peak and calculating the integrated radiative energy above the line
(see Fig. \ref{fig:margutti_lightcurve}).
We find the extra radiative energy of peak 1 to be $E_{{\rm rad,}1} \simeq 4 \times 10^{47} \erg$, and that of peak 2 to be $E_{{\rm rad,}2} \simeq 2 \times 10^{47} \erg$.
These radiated energies are somewhat smaller than the 2012a outburst but significant compared to the 2009 and 2011 outbursts (e.g., \citealt{Pastorello2012}).

The relations that hold for the orbital periods between the launching of two consecutive ejecta in our surviving companion scenario are
\begin{equation}
\begin{split}
P_{b1} &\simeq t_1 - t_b - \frac{r_1}{v_{{\rm ej},1}} + \frac{r_b}{v_{{\rm ej},b}} \\
&\simeq  40~{\rm day} - 12 \left( \frac{r_1-r_b}{1 \times 10^{15} \cm} \right)  \left( \frac{v_{\rm ej}}{10\,000 \km \s^{-1}} \right)^{-1} ~{\rm day} ,
\end{split}
\label{eq:Pb1}
\end{equation}
and
\begin{equation}
\begin{split}
P_{12} &\simeq t_2 - t_1 - \frac{r_2}{v_{{\rm ej},2}} + \frac{r_1}{v_{{\rm ej},1}} \\
&\simeq  25~{\rm day} - 1 \left( \frac{r_2-r_1}{1 \times 10^{14} \cm} \right)  \left( \frac{v_{\rm ej}}{10\,000 \km \s^{-1}} \right)^{-1} ~{\rm day}.
\end{split}
\label{eq:P12}
\end{equation}
We note that the large mass transfer process that causes the orbit to shrink occurs, according to our scenario, near the periastron passage at time $t_b-r_b/v_{\rm ej,b}$.
The mass transfer continues somewhat beyond the periastron passage (see section \ref{sec:transfer}).
This implies that the shortening of the orbit has not been completed by that time, and the orbital period $P_{b1}$ is somewhat larger than the final orbital period $P_{12}$ as calculated in equation (\ref{eq:P12}).
We take the final orbital period to be $P_{12}$, and propose that after the large mass transfer that led to the 2012b outburst, the orbital period in the binary model was $25 \pm 2$ days.
The orbital period before the large mass transfer is taken to be $\sim 38$ days, as the dominant time scale found by \cite{Marguttietal2013}.

\cite{Marguttietal2013} find the total radiated energy in the 2012b outburst to be
$E_{{\rm rad,}b}=(3.2 \pm 0.3) \times 10^{49} \erg$.
We estimate the kinetic energy of the rest of the expanding gas from the 2012b outburst at
$E_{{\rm kin,}b} \sim 2 \times 10^{49} \erg$.
This gives that the total energy released in the 2012b outburst is $E_{\rm 2012b} \sim 5 \times 10^{49}\erg$.
We consider two types of companion stars.
In the first case the companion is a main sequence (MS) star of mass $M_2 \simeq 30$--$40 ~\rm{M_\odot}$,
and a radius of $R_2 \simeq 6$--$8 ~\rm{R_\odot}$.
In the second case the companion is a Wolf-Rayet (WR) star of mass $M_2 \sim 20 ~\rm{M_\odot}$ and a radius of $R_2 \simeq 1$--$2 ~\rm{R_\odot}$.
Such WR stars are known to exist, and can evolve from a star with an initial mass of $\sim 60 ~\rm{M_\odot}$
\citep{Georgyetal2012, Sanderetal2012}.
In the second case the companion was initially more massive than the progenitor of the (present day) LBV.
It evolved beyond the main-sequence earlier, transferred mass to the now primary star, and became a WR star.
We propose MS and WR stars as possible companions for the following reasons:
(1) The star should be massive enough for its gravity to affect the LBV and trigger mass transfer.
(2) The star should have $M/R$ large enough for the accretion energy to be sufficiently large to account for
the observed energy. It therefore cannot be a giant.

The required accreted mass to supply the 2012b outburst in our scenario is
\begin{equation}
\begin{split}
M_{\rm acc} &= \frac{2 E_{\rm acc} R_2 }{G M_2} \\
            &= 5.3
\left(  \frac{M_2}{30 ~\rm{M_\odot}} \right)^{-1}
\left( \frac{R_2}{6 ~\rm{R_\odot}} \right)
\left(\frac{E_{\rm 2012b}}{5\times10^{49} \erg}\right)   ~\rm{M_\odot}. \\
\end{split}
\label{eq:Macc1}
\end{equation}
In the second case, that of a WR companion star, the accreted mass in our model is $\sim 2 ~\rm{M_\odot}$.
Relation (\ref{eq:Macc1}) is relevant both to the surviving companion scenario discussed here and the next section, and to the merger scenario to
be discussed in section \ref{sec:merger}.
We now turn to calculate the orbital evolution of the two cases in the surviving companion scenario.

% ==========================================================
\section{Shrinkage of the Orbit by Mass Transfer}
\label{sec:transfer}
% ==========================================================

Mass transfer and mass loss can change the orbital parameters of a binary system.
Mass transfer from the more massive to the less massive component (in our case, the LBV and the companion, respectively) causes
a shrinkage of the orbit, while mass loss acts to expand it.

We use the following derivation to calculate the change in the stellar and orbital parameters (e.g., \citealt{Eggleton2006}).
We define $\dot M_{l1}$ and $\dot M_{l2}$ as the rates of mass loss to infinity
from the LBV ($M_1$, $R_1$) and the companion ($M_2$, $R_2$), respectively,
and $\dot M_{\rm acc}$ as the rate of mass transferred from the primary to the companion.
The amount of accreted mass by the companion $M_{\rm acc}$ is determined from the requirement that the accretion energy of this mass should be sufficient to
account for the energy released by the 2012b outburst (see section \ref{sec:peaks}).
The rates of change of the stellar masses are consequently
\begin{equation}
\begin{split}
&\dot M_1=-\dot M_{l1}-\dot M_{\rm acc} ~~;~~
\dot M_2=-\dot M_{l2}+\dot M_{\rm acc} ~~;~~ \\
&\dot M = \dot M_1 + \dot M_2 = -\dot M_{l1} -\dot M_{l2}.
\label{eq:Mdot}
\end{split}
\end{equation}

In this section we present the orbital evolution for the surviving companion scenario. The same equations will be used in section \ref{sec:merger} where we study the merger scenario.
We assume an eccentric binary orbit with an initial orbital period of $P_0=38$ days (see section \ref{sec:peaks}), and with an initial eccentricity of $e_0$.
We also assume that mass transfer occurs when the orbital separation is below a critical value of $r_0$.
This condition determines the duration $\Delta t_{\rm acc}$ of the mass transfer and mass loss phase.
To minimize the number of parameters of the model, and due to the lack of a good prescription for the mass transfer rate,
we assume that the mass transfer rate is constant during the mass transfer phase
$\dot M_{\rm acc} = M_{\rm acc} / \Delta t_{\rm acc}$.
As it is weakly gravitationally bound, it is very likely that the envelope of the LBV swells during the mass transfer event, as a result of tidal forces exerted by the companion.
The companion passes through the swelled LBV envelope that has a radius of $\sim r_0$.
The motion of the companion star through the LBV envelope, i.e., a temporary common envelope phase, accounts for the high mass transfer rate.
The values we chose for $e_0$ are low enough to ensure that the companion does not dive too deep into the LBV envelope and
ends up in a terminal merger during the first (major) periastron passage.

The orbital separation $\textbf{\textit{r}}$ is calculated as a function of time (\citealt{Eggleton2006})
\begin{equation}
\ddot{\textbf{\textit{r}}}(t) = -\frac{GM(t)\textbf{\textit{r}}(t)}{r^3(t)} +
\dot M_{\rm acc} \left(\frac{1}{M_1(t)}-\frac{1}{M_2(t)} \right) \dot{\textbf{\textit{r}}}(t).
\label{eq:rt}
\end{equation}
On the right hand side, the first term  is the unperturbed gravitational acceleration, and
the second term is a perturbing acceleration due to mass transfer and mass loss; the mass loss terms are hidden inside $M_i(t)$.
A critical property of this expression is that the later term depends linearly on velocity, and as such its effect is larger when the binary is near periastron.

We solve the equation numerically using a Runge-Kutta-Fehlberg integration method.
The variation in eccentricity $e(t) \equiv |\textbf{\textit{e}}(t)|$ is calculated according to
\begin{equation}
GM\textbf{\textit{e}} =
\dot{\textbf{\textit{r}}}\times(\textbf{\textit{r}}\times\dot{\textbf{\textit{r}}}) -
\frac{GM\textbf{\textit{r}}}{r}.
\label{eq:e}
\end{equation}
The Keplerian energy per unit reduced mass $\varepsilon(t)$ is calculated according to
\begin{equation}
\varepsilon(t) = \frac{1}{2} \dot{r}^2(t)
- \frac{GM(t)}{r(t)},
\label{eq:energy}
\end{equation}
and then we obtain the semi-major axis
\begin{equation}
a(t) = - \frac{GM(t)}{2\varepsilon(t)},
\label{eq:a}
\end{equation}
and the orbital period
\begin{equation}
P(t)=2\pi\sqrt{\frac{a^3(t)}{GM(t)}}.
\label{eq:P}
\end{equation}

\begin{figure*}[!t]
\includegraphics[trim=0cm 0.0cm 0cm 0.0cm, clip=true, width=0.86\columnwidth]{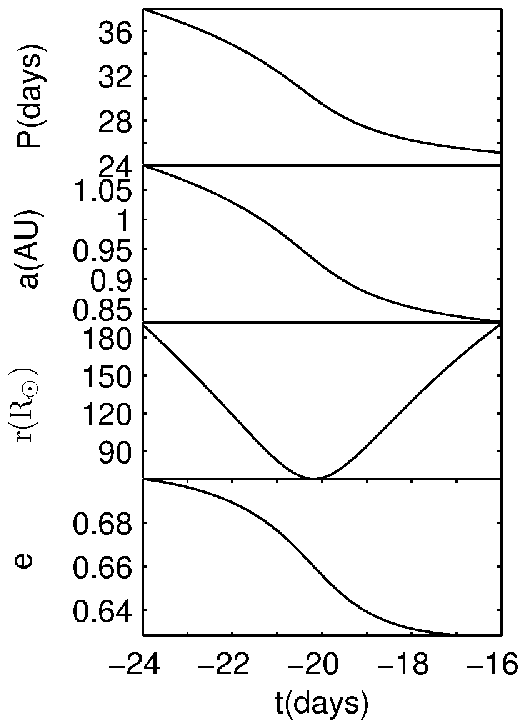}
\includegraphics[trim=0cm 0.0cm 0cm 0.0cm, clip=true, width=0.845\columnwidth]{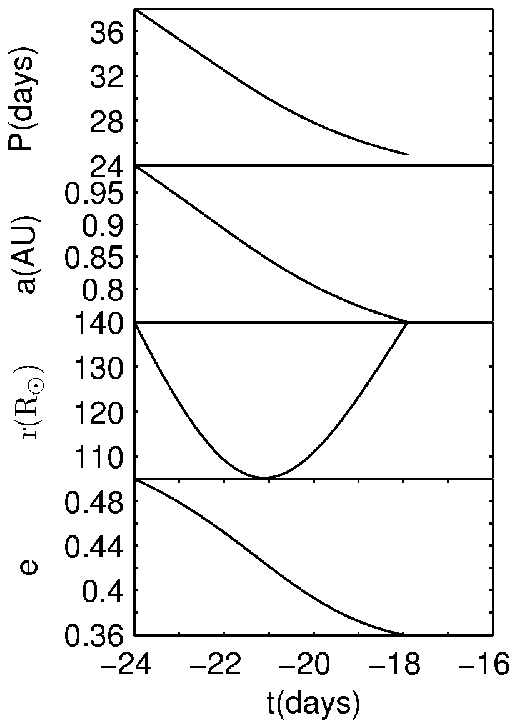}
\caption{The dynamical calculation results for model 1 with a MS companion (left) and model 8 with a WR companion (right),
both for the surviving companion scenario.
Parameters for these models are given in Table \ref{Table:comparemodels}.
The orbital period decreases from $P_f=38$ days to $P_f=25$ days as a result of mass transfer from the LBV to the suggested companion.
The time axis takes the major (first) periastron passage to have occurred at day $\sim -20$, but there is an uncertainty of $\sim \pm 2 \days$ in the exact periastron time in our scenario.
The panels are, from top to bottom: the orbital period, the semi-major axis, the binary separation and the eccentricity.
According to the suggested model the energy released by accretion onto the companion accounts for the energy of the 2012b outburst.
Note that the accretion phase duration is not the same for the different models.}
\label{fig:dynamical_calc}
\end{figure*}

We run several models for the 2012b outburst evolution varying the stellar properties as summarized in Table \ref{Table:comparemodels}.
The constraints on the evolution are that (1) the initial orbital period is $38$ days and the final one is $25$ days as discussed in section \ref{sec:peaks}, and that (2) the accreted mass onto the  star is determined by equation (\ref{eq:Macc1}).
Fig. \ref{fig:dynamical_calc} shows our model 1 -- a representative model where the companion is a MS star, and model 8 -- a representative model where the companion is a WR star.
% FFFFFFFFFFFFFFFFFFFFFFFFFFFFFFFFFFFFFFFFFFFFFF
% FFFFFFFFFFFFFFFFFFFFFFFFFFFFFFFFFFFFFFFFFFFFFF
%
\begin{table*}
\caption{
A comparison table of the parameters and the results of the models for the surviving companion scenario.
The mass transfer process continues as long as the companion was close to and within the primary surface. This is parametrized by the orbital separation $r_0$ within which mass transfer occurs.
The primary radius is not used directly in the calculation, but the model assumes that the primary swelled before the 2012b outburst to $\sim r_0$.
The calculations are constrained by an initial and final orbital periods of $P_0= 38$ days and $P_f=25$ days, respectively.
The mass transfer and loss processes occur during the time $\Delta t_{\rm acc}$ when the orbital separation obey the condition $r<r_0$.
The parameters of the model are $M_1$, $M_2$, $R_2$, $e_0$, and $r_0$ (equivalent parameter to $\Delta t_{\rm acc}$).
Other quantities are calculated.
}
\begin{center}
\begin{tabular}{l|cc|c|ccc|ccc}
\hline \hline
Parameters / Model                  &1            &2       &1A      &3       &4       &5       &6       &7       &8      \\
%                                   &             &        &        &        &        &        &        &        &       \\
\hline
Primary                             &LBV          &LBV     &LBV     &LBV     &LBV     &LBV     &LBV     &LBV     &LBV    \\
$M_1(~\rm{M_\odot})$                &80           &80      &80      &70      &70      &70      &70      &70      &70     \\
%$R_1(~\rm{R_\odot})$               &40           &40      &40      &40      &40      &40      &40      &40      &40     \\
                                    &             &        &        &        &        &        &        &        &       \\
companion                           &MS           &MS      &MS      &MS      &MS      &MS      &WR      &WR      &WR     \\
$M_2(~\rm{M_\odot})$                &40           &40      &40      &30      &30      &30      &20      &20      &20     \\
$R_2(~\rm{R_\odot})$                &8            &8       &8       &6       &6       &6       &1.5     &1.5     &1.5    \\
                                    &             &        &        &        &        &        &        &        &       \\
$r_0(~\rm{R_\odot})$                &190          &150     &180     &255     &225     &195     &210     &180     &140    \\
$\Delta t_{\rm acc}$ (days)         &8.04         &5.77    &7.36    &15.18   &13.64   &11.7    &11.2    &9.6     &6.1    \\
$M_{\rm acc}(~\rm{M_\odot})$        &5            &5       &5       &5       &5       &5       &2       &2       &2      \\
$M_1l(~\rm{M_\odot})$               &0.1          &0.1     &0.45    &0.1     &0.1     &0.1     &0.1     &0.1     &0.1    \\
$M_2l(~\rm{M_\odot})$               &0.05         &0.05    &0.05    &0.05    &0.05    &0.05    &0.05    &0.05    &0.05   \\
                                    &             &        &        &        &        &        &        &        &       \\
%$P_0$ (days)                       &38           &38      &38      &38      &38      &38      &38      &38      &38     \\
$a_0$ (AU)                          &1.09         &1.09    &1.09    &1.03    &1.03    &1.03    &0.99    &0.99    &0.99   \\
$e_0$                               &0.7          &0.6     &0.7     &0.7     &0.6     &0.5     &0.7     &0.6     &0.5    \\
%                                   &             &        &        &        &        &        &        &        &       \\
%$P_f$ (days)                       &25.14       &25.11    &25.01   &25.05   &24.98   &25.08   &24.99   &25.04   &24.96  \\
$a_f$ (AU)                          &0.83         &0.83    &0.82    &0.78    &0.78    &0.78    &0.75    &0.75    &0.75   \\
$e_f$                               &0.63         &0.49    &0.62    &0.65    &0.53    &0.39    &0.64    &0.51    &0.36   \\
\hline
\end{tabular}
\end{center}
\label{Table:comparemodels}
\end{table*}

All models assume an LBV Primary (\citealt{Berger2009}; \citealt{Smithetal2010}; \citealt{Mauerhan2013}).
Models 1, 2 and 1A assume a $80 ~\rm{M_\odot}$ LBV, and a MS companion with a mass of $40 ~\rm{M_\odot}$ and a radius $8 ~\rm{R_\odot}$.
Models 3--5 assume a $70 ~\rm{M_\odot}$ LBV, and a MS companion with a mass of $30 ~\rm{M_\odot}$ and a radius $6 ~\rm{R_\odot}$.
Models 6--8 assume a $70 ~\rm{M_\odot}$ LBV, and a WR companion with a mass of $30 ~\rm{M_\odot}$ and a radius $1.5 ~\rm{R_\odot}$.
In models 1--5 and 1A (that have a MS companion) the accreted mass is $M_{\rm acc} =5 ~\rm{M_\odot}$, and in models 6--8 (that have a WR companion) the accreted mass is $M_{\rm acc} = 2 ~\rm{M_\odot}$.
As mentioned earlier, the mass $M_{\rm acc}$ is determined such that it accounts for the energy of the 2012b outburst
by the gravitational energy released by accretion onto the companion star.
The value of the semi-major axis was determined from $P_0$ and the sum of the LBV and companion masses.
All models except model 1A have mass loss of $0.1 ~\rm{M_\odot}$ from the LBV and $0.05~\rm{M_\odot}$ from the companion.
Model 1A is the same as model 1 but with higher mass loss from the LBV, $0.45 ~\rm{M_\odot}$.

As we assume that the peaks after the 2012b outburst are a result of mass ejection close to periastron passage,
we constrain our model to have a final orbital period of $P_f=25$ days.
This constrain determines the mass transfer and loss duration $\Delta t_{\rm acc}$ (note that $r_0$ is derived from $\Delta t_{\rm acc}$
and is not a separated fitting parameter).
For each of the three sets of models ($80 ~\rm{M_\odot}$ LBV and a $40 ~\rm{M_\odot}$ MS companion; $70 ~\rm{M_\odot}$ LBV and a $30 ~\rm{M_\odot}$ MS companion;
$70 ~\rm{M_\odot}$ LBV and a $30 ~\rm{M_\odot}$ WR companion) we tried to obtain $P_f=25$ days using eccentricities of 0.7, 0.6 and 0.5.
We managed to do so for all models except for eccentricity of $0.5$ for a $80 ~\rm{M_\odot}$ LBV and a $40 ~\rm{M_\odot}$ MS companion,
where we were able to get a minimum $P_f$ of only $26$ days; we therefore omit this model from the table.
We note, however, that though we omitted this model for consistency, it does not necessarily mean that the set of parameters
used in this model cannot work. As we recall from section \ref{sec:peaks}, the final period is $25 \pm 2$ days.

As expected, we find that the higher the assumed initial eccentricity, the shorter the duration required for accretion to
reduce the orbital period to $P_f=25$ days.
For model 1A the mass loss rate was higher. As mass loss acts to increase the orbital period, this model requires a longer accretion phase
in order to reach $P_f=25$ days, compared to model 1.

We conclude that the accreted mass $M_{\rm acc}$ can result in a reduction of the orbital period from 38 to 25 days,
if an eccentric orbit is assumed.
The eccentricities are not large when one considers that the the eccentricity during the Great Eruption of
$\eta$ Carinae was $e \simeq 0.9$ (\citealt{KashiSoker2010a}). From Table \ref{Table:comparemodels} we can see
that it is possible to get this reduction in the orbital period for
a wide range of binary parameters, with no need to fine-tune the parameters.
We find our results to be encouraging for further exploration of the surviving-binary scenario for SN~2009ip.

% ==========================================================
\section{The Merger Scenario}
\label{sec:merger}
% ==========================================================

In this section we present our second scenario, the merger scenario, where we try to explain the light curve and absorption features
of the 2012b outburst by an extended binary interaction lasting two periastron passages, after which the stars merged.
We first present the motivation for the merger scenario.

\cite{Marguttietal2013} present absorption features in optical and NIR lines in a range of velocities from
$1000 \km \s^{-1}$ to $12\,500 \km \s^{-1}$.
They interpret the absorption feature as resulting from three main shells, or partial shells, that were ejected during the
2012b outburst. The absorption feature at $\sim 12\,500 \km \s^{-1}$ that first appeared on day +9 (9 days after maximum)
was attributed to the fastest shell. A second shell moving at $5500 \km \s^{-1}$ explains according to their suggestion
the absorption feature that appeared on day +28, and a third shell appeared at day +59 moving at $2500 \km \s^{-1}$.
\cite{Marguttietal2013} suggest that the first shell is asymmetric; they
found no evidence for asymmetry in the slowly-moving shells.

\begin{figure}[!t]
\includegraphics[trim=1.8cm 1.2cm 0cm 0.1cm, clip=true, width=1.1\columnwidth]{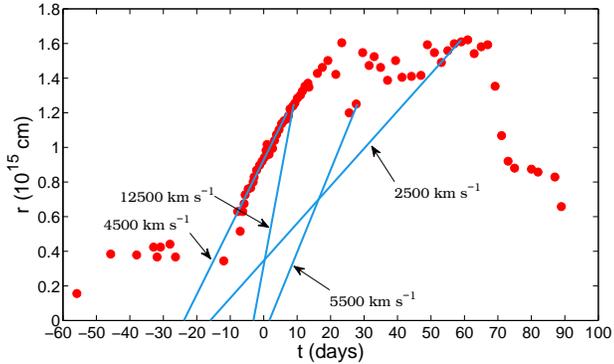}
\caption{Tracing back the main shells/clumps to their time of origin. Red points show the photospheric radius as function of time taken from
\citep{Marguttietal2013}. The  Time is measured from maximum luminosity.
The expansion velocity of the photosphere is $4500 \km \s^{-1}$ and we found it to begin at $t=-24$ days.
We find that the $2500 \km \s^{-1}$, $12\,500 \km \s^{-1}$, and $5500 \km \s^{-1}$ shells (or clumps) were ejected
at days $t=-16$, $t=-3$, and $t=+2$, respectively.
The calculation is done assuming spherical symmetry. However since some `shells' cross others, most of them seem to be large clumps or collimated outflows rather than spherical shells.
}
\label{fig:shells}
\end{figure}

We traced each shell back to its time of ejection based on its velocity and the time and photospheric radius when the
shell's absorption first appeared. We assume that the shells/clumps are not slowed down on their way to the photosphere.
The analysis is presented in Fig. \ref{fig:shells}.
It seems as the `shells' are actually clumps and/or collimated outflows, since they are crossing one another.
Only the $2500 \km \s^{-1}$ might be a complete  (or almost complete) spherical shell \citep{Marguttietal2013}.
Assuming spherical symmetry, we find that the $2500 \km \s^{-1}$, $12\,500 \km \s^{-1}$, and $5500 \km \s^{-1}$ shells/clumps were ejected
at days $t=-16$, $t=-3$, and $t=+2$, respectively.
This teaches us that the shells were not ejected simultaneously but rather during a period lasting for $\sim 20 \days$.
In addition to these main shells mentioned specifically by \cite{Marguttietal2013}, one can see in their spectra that
there are more absorption features appearing at different times and different velocities. We attribute these absorption features
to clumps that were ejected during the extended binary interaction period.
% FFFFFFFFFFFFFFFFFFFFFFFFFFFFFFFFFFFFFFFFFFFFFF
% FFFFFFFFFFFFFFFFFFFFFFFFFFFFFFFFFFFFFFFFFFFFFF

The behaviour of the expanding shells/clumps presented in Fig. \ref{fig:shells} brings us to propose an alternative to the surviving companion scenario
presented in section \ref{sec:peaks}.
In this merger scenario there were two major ejection episodes, both occurring close to periastron passages of the binary system.
The first episode was the major one. It ejected the material that caused the outburst and the photosphere expansion at
$\sim 4500 \km \s^{-1}$.
We trace the beginning of this ejection to day $\sim -24$.
It seems that the ejection of the slow $2500 \km \s^{-1}$ shell was part of this episode, possible the end of this mass ejection episode.
We take the periastron passage of the first episode to have been at day $-20$, but one should bear in mind the uncertainties of $\sim \pm 2 \days$.
Near this periastron passage our model assumes that the companion accreted mass from the LBV primary star to power the outburst, as given by equation \ref{eq:Macc1}.
The accretion caused the orbit to shrink.
We assume that the next periastron passage took place at day $\sim 0$, between the ejection times of the $5500 \km \s^{-1}$ and the $2500 \km \s^{-1}$ shells/clumps.
We take the pre-accretion binary period to be $38 \days$, and demand that the next periastron passage will be 20 days later.
Small amount of mass was accretion by the companion during second periastron passage.
However, the main shrinkage in evolution during the second periastron passage was caused by the strong dynamical interaction between the stars because
at the second periastron passage the companion is much deeper inside the primary envelope. We do not follow the evolution beyond the second periastron passage.

\begin{figure}[!t]
\includegraphics[trim=1cm 2cm 0cm 0.1cm, clip=true, width=2.10\columnwidth]{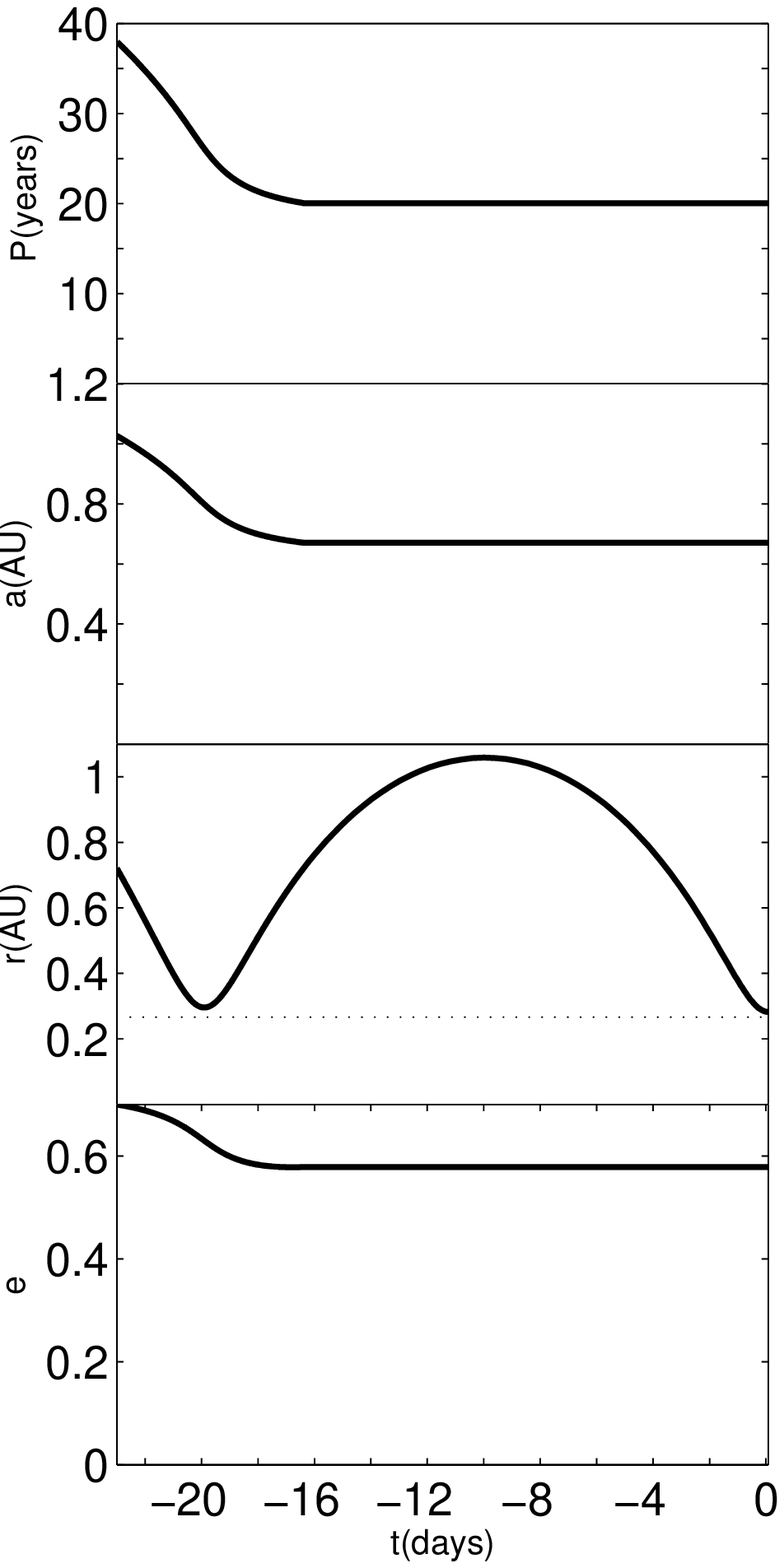}
\caption{
The suggested binary orbit for our merger scenario. Merger occurs after the second periastron passage which takes place at day $\sim 0$. 
The time axis takes the major (first) periastron passage to have occurred at days -20, but there is an uncertainty of $\sim \pm 2 \days$
in the exact periastron time in our scenario, as mass transfer starts close to periastron but not
exactly at periastron.
As a representative case we take a $M_1 = 70 ~\rm{M_\odot}$ LBV primary star and $M_2=30~\rm{M_\odot}$ MS companion in a binary system with an initial eccentricity of $e_0=0.7$.
The orbital period decreases from $P_f=38$ days as a result of mass transfer of $M_{\rm{acc}}=5~\rm{M_\odot}$.
The orbital separation below which accretion takes place, $r_0$, is fitted to $155~\rm{R_\odot}$ in order to obtain the
second periastron passage 20 days after the first (major) one.
The 20 days difference can be obtained for many sets of parameters ($e_0$, $r_0$, $M_1$, $M_2$), but here we show only one example.
The panels are, from top to bottom: the orbital period, the semi-major axis, the binary separation and the eccentricity.
}
\label{fig:two_periastron}
\end{figure}

In Fig. \ref{fig:two_periastron} we present an example for a dynamical evolution that can cause the above set of events.
We take a $M_1 = 70 ~\rm{M_\odot}$ LBV primary star and  $M_2=30~\rm{M_\odot}$ MS companion in a binary system with an initial eccentricity of $e_0=0.7$
and initial orbital period of $38 \days$.
The companion accretes $5~\rm{M_\odot}$ from the primary envelope during the first periastron passage.
The technical details of the calculation were given in section \ref{sec:transfer}.
The mass transfer reduces the orbital period such that the next periastron passage occurs $20 \days$ later.
Eccentricity is reduced to $e \simeq 0.54$.
Over all, the merger scenario can account for the different peaks and main absorption features of the 2012b outburst of SN~2009ip.
% FFFFFFFFFFFFFFFFFFFFFFFFFFFFFFFFFFFFFFFFFFFFFF
% FFFFFFFFFFFFFFFFFFFFFFFFFFFFFFFFFFFFFFFFFFFFFF

% ==========================================================
\section{Summary and Discussion}
\label{sec:summary}
% ==========================================================

The SN impostor SN~2009ip experienced a number of outbursts, most recently in September 2012 (outburst 2012b).
The several outbursts in 2009--2011 were LBV major outbursts. However, it is not yet widely agreed what was the nature of the 2012a and 2012b outbursts.
\cite{Mauerhan2013} attribute the 2012a to a supernova and the 2012b to the collision of the SN ejecta with a previously ejected gas.
This scenario was put to a question by \cite{Marguttietal2013}, who suggested that what was observed is an explosive ejection of the envelope of a massive progenitor star.
\cite{Pastorello2012} attributed the outbursts in 2012 to core instabilities that did not destroy the star.
We attribute {\it all} the outbursts to periatron passages of an eccentric binary system.
We find our interpretation to be supported by the thorough analysis performed recently by \cite{Marguttietal2013},
who found a dominant time scale of $\sim 38 \days$ in the light curve previous to the 2012b outburst.
In \cite{SokerKashi2013} the much more energetic 2012b outburst was attributed to a final merger process that occurred during a single periastron passage that powers the 2012b outburst.

In the present paper we consider the possibility that the companion did survive the major periastron passage that powered the 2012b outburst.
During the first (major) periastron passage the companion accreted $\sim 2$--$5 ~\rm{M_\odot}$ from the LBV envelope.
The liberated gravitational energy of the accreted mass can account for the total 2012b outburst energy (equation \ref{eq:Macc1}).
In the declining light curve of the 2012b outburst there are two large peaks.
The extra radiating energy in each of these two peaks is similar to the 2009--2011 outbursts (section \ref{sec:peaks}).
If we interpret the peaks to result from mass ejected during later periastron passages, then the inferred orbital period after the large mass accretion is
$\sim 25 \days$ (see equation \ref{eq:P12}). We therefore propose a scenario where the companion survived the interaction and it still exists today.
This is the surviving companion scenario discussed in sections \ref{sec:peaks} and \ref{sec:transfer}.
We perform a dynamical integration of the binary orbit, as presented for two cases in Fig. \ref {fig:dynamical_calc}, and found that a shrinkage of the orbit
from 38 to 25 days is obtainable for a rich variety of binary parameters (Table \ref{Table:comparemodels}).
If this interpretation is correct, then it is quite likely that SN~2009ip will suffer another outburst sometime in the future,
as a result of another periastron passage.
We predict that in such a case a periodicity of $\sim 25 \days$ might show up in the light curve.

The dominant time scale might be related to the orbital period of the proposed binary system before the 2012b outburst, but not necessarily.
Binary periodicities that are detected by eclipses, reflection and radial velocity have well defined periodicity.
In the case of the outbursts of SN~2009ip photometric variations provide us information
about the binary interaction.
In our model it is mainly mass transfer that is behind the photometric variations.
The energy output from mass transfer depends on the distance of the companion from the primary that varies with the orbital motion in an eccentric orbit, and the state of the unstable primary star.
The variations in the unstable primary star, e.g., in its radius, are not synchronized with the orbital period.
Hence, the primary star modulates the strength of the binary interaction,
like mass transfer, around the fundamental binary period.
Departure from spherical symmetry might also contribute to unstable periodicity,
e.g., jets-CSM interaction \citep{TsebrenkoSoker2013} and/or jets' precession. 
We therefore do not expect the photometry to be directly correlated with binary periodicity.
As we showed above, our model is not sensitive to the parameters, and this is true also for the initial orbital period.
It is possible that the dominant time scale found by \cite{Marguttietal2013} does not reflect the orbital period.
In that case our suggested orbital period of $38 \days$ before the 2012b outburst would be regarded as an assumption of the binary model.

Multi-band analysis of the light curve of across the 2012b event hints on other periods of 24, 16, and 12 days (\citealt{Martinetal2013a}; \citealt{Martinetal2013b}).
\citealt{Martinetal2013b} suggested that the fluctuations are related to a single pulsating star
which was not destroyed in the 2012 eruptions.
As the photosphere does not expand after day 20 post-maximum light, it is not clear how collision will affect the photospheric emission.
Instead, in our proposed scenario the fluctuations might result from the clumpy structure of the ejected shells, and from the relaxing LBV star.
The shells in our model are bipolar outflows (jets), that might be clumpy.
Large clumps can cause fluctuations in the light curve. For example, a large dense clump might move at a lower than the shell speed,
and collide at a later time with the gas near the photosphere.

We also considered the merger scenario, where the companion merged with LBV primary in the second periastron passage (section \ref{sec:merger}).
This is motivated by the ejection times of the three main shells/clumps and the photosphere discussed by \cite{Marguttietal2013}, as we presented in Fig. \ref{fig:shells}.
The ejection of the clumps and shells occurred at the two periastron passages, around times $-20$ and $\sim 0$ days.
The amount of accreted mass in the first (major) periastron passage is the same as in the surviving companion scenario, but it was concentrated closer to
periastron, and hence orbital shrinkage was larger and the second periastron passage occurred only $\sim 20 \days$ after the major one (Fig. \ref{fig:two_periastron}.

Although we concentrated on four ejected shells/clumps, the merger scenario can account for most other absorbing features.
The collisions of collimated outflows with previously ejected shells occurring in optically thick medium can be unstable and lead to the formation
of many small clumps moving at different velocities \citep{AkashiSoker2013}. The less dense parts of each ejection clumps will be
slowed down and may lead to formation of other clumps.
This will be studied in the future with 3D hydrodynamical simulations.

In both our proposed scenario the LBV star suffered a traumatic disturbance from the companion during the 2012b major outburst (first periastron passage) and must relax.
The dynamical time of the LBV star is $\sim 2$ days, so dynamical relaxation will last for a few days.
Thermal relaxation is longer even.
The rotating LBV star might as well have magnetic activity. These activities might also cause fluctuations in the declining light curve of the 2012b outburst.
We point out that the light curve of V838~Mon (that was compared to SN~2009ip in \citealt{SokerKashi2013}) had fluctuations around the main three peaks \citep{Tylenda2005}.
These can be explained by clumpy ejected shells and clumps formed in the shells interaction, as we propose for SN~2009ip.

The 24 days period found by \cite{Martinetal2013b} may be the same as the $\sim 25$ days binary post-2012b outburst period we propose here in the surviving companion scenario.
\cite{Martinetal2013b} included the small peaks in addition to the two large peaks (1 and 2) in their analysis of the 2012b outburst light curve and therefore the
center of this wide peak in the power spectrum is shifted to 23 days.

The light curve of the pre-explosion outburst and the explosion of SN 2010mc \citep{Ofek2013}, a type-IIn SN, is very similar to the light curve of the 2012a and 2012b
outbursts of SN~2009ip (\citealt{Smithetal2010}; \citealt{Marguttietal2013}).
\cite{Soker2013} suggested that the pre-explosion outburst of SN 2010mc was powered by a binary interaction.
The similarity of the light curves may supports the opinion that the 2012b outburst of SN~2009ip was also a SN explosion.
Or, more speculatively, that SN2010mc is an impostor and its explosion was not a type-IIn SN.
The 2000 and 2008 outbursts of the LBV in NGC 3432 are another example of an LBV whose outbursts may have been triggered by a WR
companion (\citealt{Pastorello2010}; \citealt{Kashisoker2010b}).
Together with the companion triggered outbursts of $\eta$ Carinae and P-Cygni \citep{Kashi2010}, it seems as companions play
a significant role in setting the conditions for giant LBV eruptions to occur.

SN~2009ip belongs to growing group of transient events that are less energetic than normal supernovae.
We take the view that most of these events are powered by mass transfer or merger (that is an extreme case of mass transfer)
between two binary components merger (\citealt{Kashisoker2010b}\footnote[2]{see also \url{http://physics.technion.ac.il/~ILOT/}}).
The interaction between the binary companion and the LBV may create an accretion disk or belt around the companion and maybe
blow jets that will create a bipolar structure.
\cite{SokerKashi2013} predicted that the ejecta of the 2012b outburst will have a bipolar, or even a more complicated structure.
The asymmetry of the outburst was later observed by \cite{Marguttietal2013}.
A bipolar outburst is consistent with both the scenarios we propose here.

\section*{Acknowledgments}
We thank Roger Chevalier, Raffaella Margutti, John C. Martin, Andrea Pastorello, Jose L. Prieto and an anonymous referee for helpful comments.
AK acknowledges support from NASA under ATP grant NNX11AI96G, and from the National Science Foundation under Grant No. NSF~ PHY11-25915.

%--------------------------
{}

\end{document}